\documentclass[aps,pra,twocolumn,showpacs,preprintnumbers,amsmath,amssymb,superscriptaddress]{revtex4-1}

\usepackage{graphics}
\usepackage{graphicx}
\usepackage{epsfig}
\usepackage{amsmath}

\newcommand{\beq}{\begin{equation}}
\newcommand{\eeq}{\end{equation}}
\newcommand{\bea}{\begin{eqnarray}}
\newcommand{\eea}{\end{eqnarray}}
\usepackage{dcolumn}
\usepackage{bm}
\def\Rb87{^{87}\rm{Rb}}                 
\def\Li6{^{6}\rm{Li}}                   

\def\ex{{\mathbf e}_x}                            
\def\ey{{\mathbf e}_y}                            
\def\ez{{\mathbf e}_z}                            
\DeclareMathAlphabet\mathbfcal{OMS}{cmsy}{b}{n}

\begin{document}
\title{Dynamically slowed collapse of a Bose-Einstein condensate with attractive interactions}
\author{R.~L. Compton}
\affiliation{Joint Quantum Institute, National Institute of Standards and Technology, and University of Maryland, Gaithersburg, Maryland, 20899, USA}
\author{Y.-J. Lin}
\affiliation{Joint Quantum Institute, National Institute of Standards and Technology, and University of Maryland, Gaithersburg, Maryland, 20899, USA}
\author{K. Jim\'{e}nez-Garc\'{i}a}
\affiliation{Joint Quantum Institute, National Institute of Standards and Technology, and University of Maryland, Gaithersburg, Maryland, 20899, USA}
\affiliation{Departamento de F\'{\i}sica, Centro de Investigaci\'{o}n y Estudios Avanzados del Instituto Polit\'{e}cnico Nacional, M\'{e}xico D.F., 07360, M\'{e}xico}
\author{J.~V. Porto}
\affiliation{Joint Quantum Institute, National Institute of Standards and Technology, and University of Maryland, Gaithersburg, Maryland, 20899, USA}
\author{I.~B. Spielman}
\affiliation{Joint Quantum Institute, National Institute of Standards and Technology, and University of Maryland, Gaithersburg, Maryland, 20899, USA}
\email{ian.spielman@nist.gov}

\begin{abstract}
We rapidly change the scattering length $a_s$ of a $^{87}$Rb Bose-Einstein condensate by means of a Feshbach resonance, simultaneously releasing the condensate from its harmonic trapping potential.  When $a_s$ is changed from positive to negative, the subsequent collapse of the condensate is stabilized by the kinetic energy imparted during the release, resulting in a deceleration of the loss rate near the resonance.  We also observe an increase in the Thomas-Fermi radius, near the resonance, that cannot be understood in terms of a simple scaling model.  Instead, we describe this behavior using the Gross-Pitaevskii equation, including three-body recombination, and hypothesize that the increase in cloud radius is due to self-interference of the condensate resulting in the formation of concentric shells.
\end{abstract}

\pacs{75.75.+a,75.40.Gb}

\maketitle

\section{\label{sec:Intro}Introduction}
Bose-Einstein condensation of neutral atoms is usually realized in systems with positive atomic scattering length $a_s$; the resulting repulsive interaction allows the formation of large $\approx 10^6$~atom condensates in harmonic potentials.  Even for weakly attractive interactions, however, the zero point kinetic energy of the trap can stabilize quantum degenerate gases against collapse at sufficiently low density \cite{Bradley:PRL1997,Ruprecht:PRA1995,Gerton:Nature2000,Altin:PRA2011}.  Strongly attractive condensates have been produced in an important class of experiments that uses Feshbach resonances to rapidly tune the scattering length $a_s$ from positive to negative, but are dramatically unstable, resulting in collapsing clouds that expel atoms in bursts \cite{Roberts:PRL2001, Duine:PRL2001, Donley:Nature2001}.  These attractive condensates were formed at low density in weak harmonic traps, thereby avoiding strong three body recombination that scales as density cubed.  Here we investigate the stability of an \textit{untrapped} $^{87}$Rb Bose-Einstein condensate tuned to negative scattering length in the vicinity of a Feshbach resonance.  

Near a Feshbach resonance the scattering length tracks a time dependent magnetic field $B(t)$ as
\begin{equation}
    a_s(t)=a_{\rm bg}\left[1-\frac{\Delta B}{B(t)-B_0}\right],
\label{eq:scattering}
\end{equation}
where $a_{\rm bg}$ is the background scattering length ($\approx 5.3$~nm for $^{87}$Rb); and $\Delta B$ and $B_0$ are the width and center of the resonance, respectively.  For a model resonance width $\Delta B_{\rm th} = 17\ \mu$T \cite{Marte:PRL2002}, this results in a zero crossing for $a_s$ at $B-B_0=17\ \mu$T, as shown in Fig.~\ref{fig:numandwidth}(a).  Near this zero crossing, cloud density increases sharply, resulting in rapid three-body losses.  In our experiment, we reduce density and minimize three body recombination by releasing the atoms from the harmonic trap before rapidly ramping $B(t)$ to a final field $B_f$ close to $B_0$, as done by Volz \textit{et al.} \cite{Volz:PRA2003}.  On the high field side of the resonance ($B_f - B_0 > \Delta B_{\rm th}$), Volz \textit{et al.} observed the expected decrease in the condensate's Thomas-Fermi radius $R_{\rm TF}$ as a function of decreasing $B_f$, corresponding to a decrease in positive $a_s$.  For final field settings below the expected zero crossing for $a_s$, however, they observed an increase in $r_{\rm TF}$, relative to the minimum cloud width observed near the expected zero crossing.  They interpreted the swelling in $r_{\rm TF}$ as arising from a destabilization of the condensate in the attractive regime, similar to the instability responsible for the explosive condensates of Refs.~\onlinecite{Roberts:PRL2001, Duine:PRL2001, Donley:Nature2001}.

In this work, we observe a similar elevation in $r_{\rm TF}$ of an $^{87}$Rb condensate, but only for $0<B_f-B_0\le 11(1)~\mu$T, 35\% lower than the expected crossover to negative scattering length at $B_f-B_0 = \Delta B_{\rm th} = 17~\mu$T \cite{Marte:PRL2002}.  In contrast to the interpretation of Ref.~\onlinecite{Volz:PRA2003}, we believe that the observed increase in cloud width, rather than indicating instability or explosion of the condensate, is related to the sudden release of the condensate from the trap, which happens much more quickly than the $\approx$1~ms field ramp to negative scattering length.  There is a short period of time after release, therefore, when the mean field energy of the condensate is still positive.  This positive mean field energy is converted to kinetic energy during the brief time after release before the scattering length becomes negative.  In analogy to the the stabilization of a weakly attractive condensate by its zero-point kinetic energy in a harmonic trap, we suggest that a similar stabilization arises from the conversion of mean field to kinetic energy prior to the reversal of the mean field from repulsive to attractive.  In the untrapped case, however, the dynamics are more complex, since the condensate, at the moment of release, becomes an outward traveling, approximately spherical matter wave.  Upon reversal of the scattering length, it then becomes a complicated superposition of inward traveling and outward traveling spherical waves that interfere.  (The description in terms of interfering matter waves provides a correct, although not strictly needed intuition for this effect: it can also be fully explained using a suitable set of classical fluid hydrodynamic equations, including 3-body recombination.) While we do not directly observe interference phenomena in this experiment, we show that aspects of this hypothesis can be successfully modeled using a combination of analytical and numerical arguments.

%
\section{\label{sec:Exp}Experimental Considerations}
We prepare a Bose-Einstein condensate (BEC) of $N \approx 2\times10^6$ atoms in a crossed optical dipole trap \cite{Lin:PRA2009}, formed by a pair of 1064~nm laser beams crossing in the $\ex-\ey$ plane, with final harmonic trapping frequencies of $\{\omega_x,\omega_y,\omega_z\}/2\pi=\{70,55,73\}$~Hz.  Our $|f=1,m_F=-1\rangle$ BEC starts in a small $B\approx0.1$~mT bias field along $\ez$ (vertical) before we transfer the atoms to $|f=1,m_F=+1\rangle$ by radio-frequency adiabatic rapid passage.  Stern-Gerlach separation of the spin states reveals no visible population outside $|f=1,m_F=+1\rangle$.)  Following the transfer, the bias field is set to $B\approx 100.7$~mT, near the Feshbach resonance under study.

Because the widest $^{87}$Rb Feshbach resonance, centered at $B_0 = 100.7$~mT, has a theoretical width $\Delta B_{\rm th}$ of only $17~\mu$T (170~mG, see Refs.~\onlinecite{Marte:PRL2002} and \onlinecite{Volz:PRA2003}), effective tunability of the scattering length requires magnetic field resolution and stability on the level of 10~ppm, which we achieve.  Our ``Feshbach coils" are mounted in a Helmholtz configuration, and each consists of 48 turns of hollow copper tubing. Four sets of inlets and outlets provide approximately 5 l/min of cooling water to each coil.  A $300$~A current generates the required bias field, which settles to the desired set-point in less than 1~s \cite{Current}.  An additional set of four-turn trim coils can rapidly ($\approx 1$~ms) tune the field within $\pm0.2$~mT (2~G) of the resonance.  When current is switched into the Feshbach coils, the initial field response overshoots the resonance, but at a high slew rate so that few atoms are lost.  Since some losses and heating are unavoidable, we allow an additional 2~s of free evaporation after the current stabilizes, resulting in a nearly pure BEC of $\approx~500\times10^3$~atoms, at a field that is either $\approx 100.54$~mT or $\approx 100.94$~mT.

\section{\label{sec:Results}Results and Discussion}
\subsection{\label{sec:Feshbach}Feshbach Resonance}
%
\begin{figure}
\center
\includegraphics{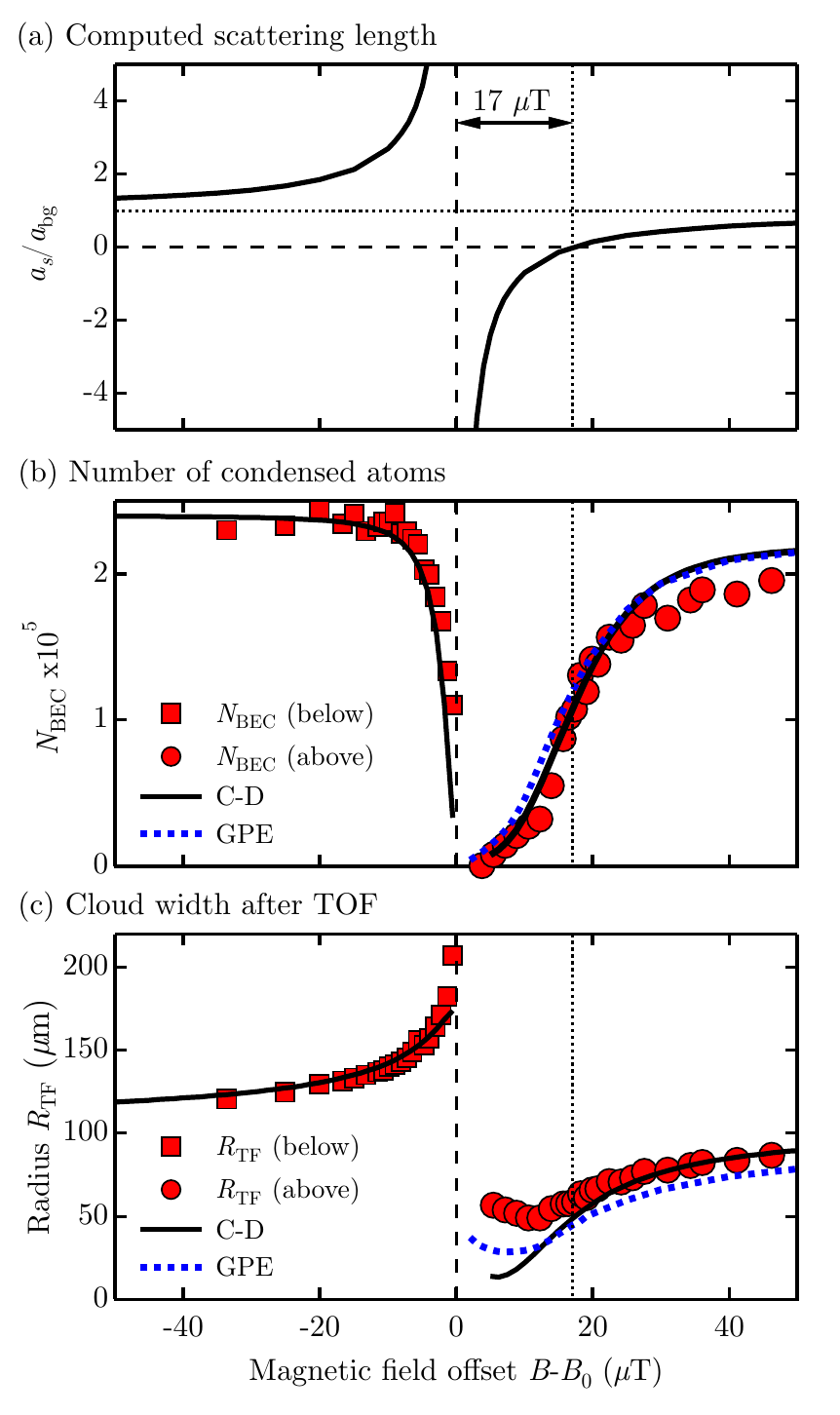}
\caption{
(a) Computed scattering length in the vicinity of a Feshbach resonance (solid curve) with width $\Delta B  = 17 \mu$T, from Eq.~\eqref{eq:scattering}.  Far from resonance, $a_s/a_{\rm bg}$ approaches unity.  (b) Number of condensed atoms versus magnetic field revealing a sharp loss feature near the resonance field $B_0$.  Data are shown for measurements that approach $B_0$ from below (squares) and from above (circles).  (c) Cloud width after TOF (symbols) is modified by the increasing (decreasing) scattering length below (above) the resonance.  Zero free parameter models of the data in (b) and (c) are based on both Castin-Dum scaling (solid curves), and 1D GPE simulations (dotted curves), as described in the text.
}
\label{fig:numandwidth}
\end{figure}

For the data displayed in Fig.~\ref{fig:numandwidth}, the BEC was first prepared at a field either slightly above or below the Feshbach resonance, and the atom trap was then turned off and the trim coils were simultaneously ramped by $\approx100\ \mu{\rm T}$ to a final field nearer the center of the resonance.  The transient field from this 500~$\mu$s ramp couples inductively to the 48 turn Feshbach coils and is sufficient to perturb the power supply on a level of $\approx100$~ppm.  We therefore actively shield the commercial power supply using a second set of trim coils, which have a greater inductive coupling to the Feshbach coils but a lesser contribution to the total field at the location of the atoms.  The measured transient field response associated with a step change of the trim coils in the presence of the shielding system has an exponential time constant of approximately $3.6$~ms.  But because of the active shielding, the amplitude of the transient is much reduced.  Following a 0.2~mT (2~G) step change, we are therefore able to stabilize the magnetic field to within $1~\mu$T (10~mG) on a timescale of $\approx 1$~ms.  This stability is maintained for longer than 10~s following the final trim coil adjustment.  Following release of the trap, and the final 500~$\mu$s field ramp, the condensate is allowed to freely evolve for 20.0~ms under the influence of the Feshbach resonance.  The field is then rapidly lowered, almost to zero, and the cloud expands for another 9.6~ms.  This final expansion increases the cloud size and reduces the optical depth, facilitating absorption imaging (along $\ez$) to determine the 2D column density after 30.1~ms total time-of-flight (TOF).  We fit the 2D image to the sum of Thomas-Fermi and Gaussian distributions to obtain condensate and thermal characteristics.  

Figure~\ref{fig:numandwidth} shows the number of Bose-condensed atoms $N_{\rm BEC}$ in the condensate along with the Thomas-Fermi radius $R_{\rm TF}$ as a function of the final field setpoint $B_f$, relative to the center of the resonance $B_0$.  Accelerated 3-body recombination near the resonance gives rise to the sharp loss feature in atom number in Fig.~\ref{fig:numandwidth}(b).  The loss feature is significantly sharper for measurements on the low field side (squares) than on the high field side (circles).  This is expected, since three-body recombination scales as $\rho^3$, and much higher cloud densities are expected on the high field side where $a_s$ is only weakly repulsive or even attractive.  The loss feature therefore extends to higher field offset on the high field side.  Somewhat surprising, however, is the change in curvature of the loss feature on the high field side.  We naively expect the curvature of the atom loss curve to remain negative until no condensed atoms remain in the cloud.  Instead, we see a sharp inflection point at $B_f-B_0\approx12~\mu$T, below which the slope of the remaining atom number versus final field setpoint decreases significantly.  This inflection is a general feature observed over a range of experimental protocols and suggests a stabilizing influence on the condensate in a region where instability (either rapid collapse or explosion) might have been expected.  This stabilizing influence is aided in part by the losses themselves, which decrease the magnitude of the attractive mean field energy.

The Thomas-Fermi radius $R_{\rm TF}$ [squares in Fig.~\ref{fig:numandwidth}(c)] increases dramatically on the low field side of the resonance by up to a factor of 2 relative to its background size of $\approx 100$~$\mu$m.  Although we observed much larger cloud widths for fields even closer to the resonance, the clouds lose their bimodal appearance, and lack the expected aspect ratio given the initial trap anisotropy \cite{Castin:PRL1996}.  These data points are therefore excluded from Fig.~\ref{fig:numandwidth}.  We rely primarily on the low field divergence of $R_{\rm TF}$ to identify the center of the resonance, for which radio-frequency spectroscopy gives $B_0=100.742(1)$~mT, consistent with previous measurements \cite{Volz:PRA2003}.

Approaching the resonance from the high field side, the cloud width [filled circles in Fig.~\ref{fig:numandwidth}(c)] decreases as $a_s$ passes through 0 and becomes increasingly negative.  We observe a minimum in cloud width at a field offset of $B-B_0 = 11(1)~\mu$T, almost half the value of 20(3)~$\mu$T obtained in Ref.~\onlinecite{Volz:PRA2003}, but comparable to the inflection point in atom loss discussed above.  We find that the position of the field minimum is largely independent of trapping frequencies and other experimental parameters such as changes to timing protocol.  The disagreement between our observations and those of Ref.~\onlinecite{Volz:PRA2003} is therefore a mystery.  Comparison to Gross-Pitaevskii equation (GPE) simulations (discussed below) implies that the position of this minimum occurs several $\mu$T below the zero crossing of $a_s$, so that it cannot be interpreted as a direct measure of resonance width.

%
\subsection{\label{sec:K3}Three Body Losses}

\begin{figure}
\center
\includegraphics{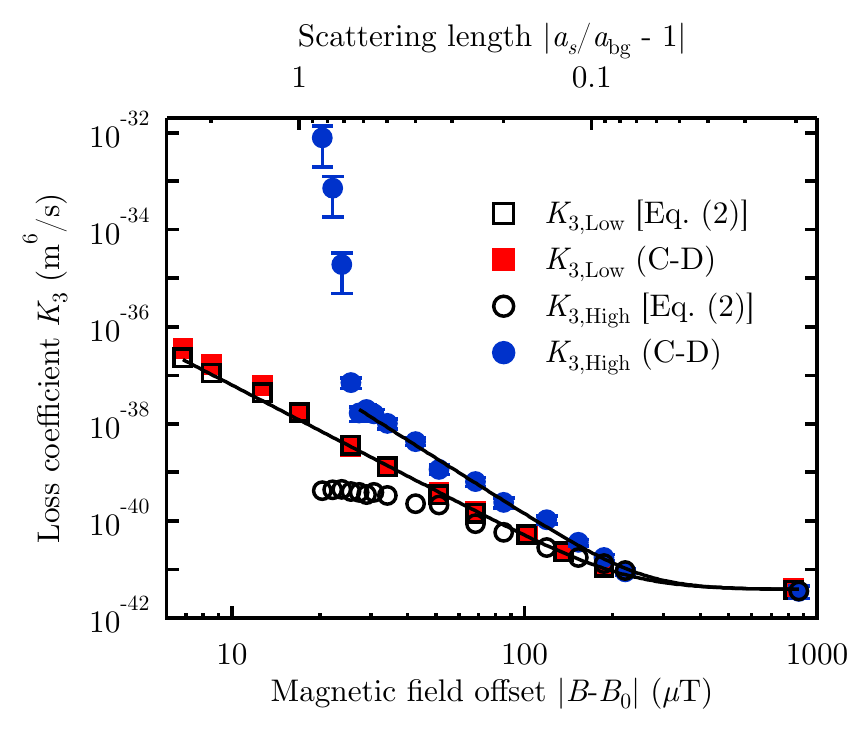}
\caption{
The loss rate coefficient $K_3$ can be extracted from atom number versus holding time data using the analytical model of Eq.~\eqref{eq:K1andK3}, with the prefactor of Eq.~\eqref{eq:alpha}, neglecting the settling time for $a_s(t)$.  Below the resonance (open squares), density is relatively low, and the results of this procedure are nearly indistinguishable from a numerical model based on the Castin-Dum scaling parameters (closed squares).  Uncertainties, based on one-sigma uncertainties from the fits to Eq.~\eqref{eq:K1andK3}, are smaller than the symbols.  The fit (solid curve) to a power law $K_3 \propto |B_f-B_0|^\beta$ yields $\beta_{\rm low} = 3.1$.   Above the resonance, the results of the fits to Eq.~\eqref{eq:K1andK3} (open circles) differ from the results of the Castin-Dum model (closed circles).  For data points below $25~\mu$T, losses occur on a timescale of a few ms, faster than the inverse trap frequency, and even the assumptions of the Castin-Dum numerical model break down.  A power law fit (solid curve) to the high field Castin-Dum data for $B_f-B_0>25~\mu$T yields $\beta_{\rm high}=$3.8.
}
\label{fig:alphaK3}
\end{figure}

Quantitative analysis of atom loss and cloud width data requires knowledge of the field dependence of the 3-body loss rate constant $K_3$ in the vicinity of the resonance.  Three-body recombination occurs when two atoms associate into a molecular state that is deeply bound relative to the $\approx$100~nK temperature of the partially condensed cloud.  Conservation laws require a third atom that gains kinetic energy; all three atoms gain sufficient momentum to depart the cloud.  Three-body losses increase as $K_3 \rho^3$, where $K_3$ is the 3-body loss rate constant and $\rho$ is the local density of atoms within the cloud.  As $a_s$ decreases from positive to negative, the cloud density increases, accelerating losses.  Also, in the vicinity of a Feshbach resonance, the increased overlap with the molecular state causes $K_3$ itself to increase dramatically, further increasing the loss rate.

$K_3$ was obtained by measuring the remaining atom number $N$ after holding the atoms \textit{in the trap} at a final field $B_f$ for a variable length of time $t$.  These data were analyzed in two different ways.  First, we fit the resulting decay curves to an analytical model
\begin{equation}
    \frac{dN}{dt}=-K_1 N -K_3\int d^3r n^3 = -K_1 N - \alpha K_3 N^{9/5},
\label{eq:K1andK3}
\end{equation}
where $K_1$ is the one-body loss rate constant and $n$ is the density.  Two-body losses are assumed to be small \cite{Soding:ApplPhysB1999}.  The prefactor $\alpha$ results from the evaluation of the integral with the assumption that the BEC retains a Thomas-Fermi profile during its evolution, and that $a_s$ reaches steady state on a time scale that is short compared to the losses:
\begin{equation}
\alpha = \frac{5^{4/5}m^{12/5}\omega^{12/5}}{56(3^{1/5})a_s^{6/5}\hbar^{12/5}\pi^2}.
\label{eq:alpha}
\end{equation}

Figure~\ref{fig:alphaK3} shows the field dependence of $K_3$, obtained by this analytical model, both below (open squares) and above (open circles) the resonance.  On the low-field side of the resonance (squares), $K_3$ exhibits a simple power law dependence $K_3 \propto |B_f-B_0|^\beta$ with a scaling exponent $\beta_{\rm low}= 3.1(1)$, slightly higher than the value of approximately 2.6 that can be extracted from the data of Ref.~\onlinecite{Smirne:PRA2007}.

In addition to the analytical model of Eqs.~\eqref{eq:K1andK3} and \eqref{eq:alpha}, we also fit the data of Fig.~\ref{fig:alphaK3} to a numerical model based on a scaling law \cite{Castin:PRL1996}, as demonstrated in Ref.~\onlinecite{Volz:PRA2003}, with two important extensions.  First, our numerical solution of the Castin-Dum scaling equations incorporates the time dependence of $a_s(t)$ from Eq.~\eqref{eq:scattering}, using the measured rise-time of the field $B(t)$ and the predicted field width of the resonance  $\Delta B_{\rm th} = 17~\mu$T~\cite{Marte:PRL2002}.  The time dependence of $B(t)$ includes a $500~\mu$s ramp to within $1~\mu$T of the final value, followed by an exponential decay to the final setpoint, with a time constant of 3.6 ms, as determined by RF spectroscopy.

Our second extension to the analysis of Ref.~\onlinecite{Volz:PRA2003} is the self-consistent inclusion of realistic time-dependent losses in atom number.  We assume a power law dependence $K_3 = \alpha |B-B_0|^\beta$, fit the loss data to the Castin-Dum scaling equations based on initial guesses for $\alpha$ and $\beta$, and obtain a new set of values for $K_3$ vs $B$, from which new values for $\alpha$ and $\beta$ are obtained.  We iterate this procedure until it converges on a stable loss exponent.  In the case of the low field data, this numerical procedure yields results (closed squares in Fig.~\ref{fig:alphaK3}) that are equivalent to those obtained from the fits to Eq.~\eqref{eq:K1andK3}.

For the high field data, however, there is a striking difference between the results of the analytical model~[Eq.~\eqref{eq:K1andK3}] (open circles) and the Castin-Dum model (closed circles).  This is because the time dependence $a_s(t)$, neglected in Eq.~\eqref{eq:K1andK3}, is more important on the high field side, where decreasing $a_s$ shrinks the Thomas-Fermi radius of the cloud, increasing density, and accelerating losses.  For values of $|B_f -B_0|>25~ \mu$T, the Castin-Dum model yields results for $K_3$ that appear to obey a power law with $\beta_{\rm high} = 3.8(1)$, significantly higher than $\beta_{\rm low}$.  Using Eq.~\eqref{eq:scattering}, we note that $K_3$ also obeys a power law with $|a/a_{\rm bg}-1|$, which mirrors the abscissa of Fig.~\ref{fig:alphaK3} (and does not scale with $|a/a_{\rm bg}|$ directly).  The universal scaling law $K_3 \propto a^4$ \cite{Fedichev:PRL1996}, which has been confirmed for $a/a_{\rm bg} \gg 1$ \cite{Weber:PRL2003}, is therefore not obeyed for our data.  However, this universality law is not expected to to hold for small $a$, or even very close to the resonance \cite{Stenger:PRL1999}, and we are therefore not necessarily surprised to find $\beta_{\rm high} \ne \beta_{\rm low}$.

For the Castin-Dum analysis of the high side data with $|B_f-B_0|<25~\mu$T, $K_3$ appears to diverge rapidly.  This is most likely a misleading result of the departure of the cloud density from an ideal Thomas-Fermi profile, as the loss rate approaches the $\approx 1/\omega$ equilibrium timescale for the cloud, where $\omega=(\omega_x \omega_y \omega_z)^{1/3}$ is the geometric average trapping frequency.  Fits to the loss data in this regime are visibly poor and yield large uncertainties.  Despite the probable breakdown of the Castin-Dum model very close to resonance, we believe that for $|B_f-B_0|>25~\mu$T it provides a better measure of $K_3$ than the fits to Eq.~\eqref{eq:K1andK3}, due to the inclusion of the time dependence of $a_s(t)$.  For subsequent analysis of the TOF data of Fig.~\ref{fig:numandwidth}, we have therefore assumed $K_3 \propto (B-B_0)^{3.8}$, as obtained from the Castin-Dum model, and have further assumed that this power law holds even for $B-B_0<25~\mu$T, with constant $\beta_{\rm high}$.

\subsection{\label{sec:C-D}Castin-Dum Analysis}
With the foregoing assumptions, we now attempt to fit the \textit{untrapped} behavior of Fig.~\ref{fig:numandwidth} using the same Castin-Dum scaling argument that was just used to analyze the \textit{trapped} loss data of Fig.~\ref{fig:alphaK3}.  A zero free parameter theory curve (solid curve), using the values for $\beta_{\rm low}$ and $\beta_{\rm high}$ obtained above, reproduces the loss feature of Fig.~\ref{fig:numandwidth}(b) quite well, including the inflection in the slope of the loss curve on the high field side.  Based on this model, we understand this reduction in loss rate in terms of a reduction of the mean field energy that drives the collapse, since mean field energy depends on atom number.

The Castin-Dum theory curve of Fig.~\ref{fig:numandwidth}(c) likewise yields reasonable agreement for $B_f<B_0$, where the maximum cloud width (excluding points closer to $B_0$ that appear to have a thermal profile) corresponds to a factor of 35 increase in $a_s$.  On the high field side of the resonance, the Castin-Dum model once again produces an inflection in the slope of $R_{\rm TF}$ vs $B_f-B_0$.  As noted previously, rapid losses reduce the negative mean field energy that drives the collapse, thereby slowing the collapse.  However, the Castin-Dum model has $R_{\rm TF} \rightarrow 0$ without an upturn at very low fields, in disagreement with the experimentally observed minimum and subsequent increase in cloud width with decreasing field offset.  This departure from experimental results is consistent with the presumably artificial increase in $K_3$ obtained from application of the Castin-Dum model to the loss data of Fig.~\ref{fig:alphaK3} and, as discussed above, is likely due to the departure of the ultra-cold cloud from a simple Thomas-Fermi profile.  Instead, to reproduce the observed increase in cloud width, which has previously been interpreted as an instability \cite{Volz:PRA2003}, we must model the system within the Gross-Pitaevskii equation (GPE).

%
\subsection{\label{sec:GPE}Gross-Pitaevskii Simulation}
\begin{figure}
\center
\includegraphics{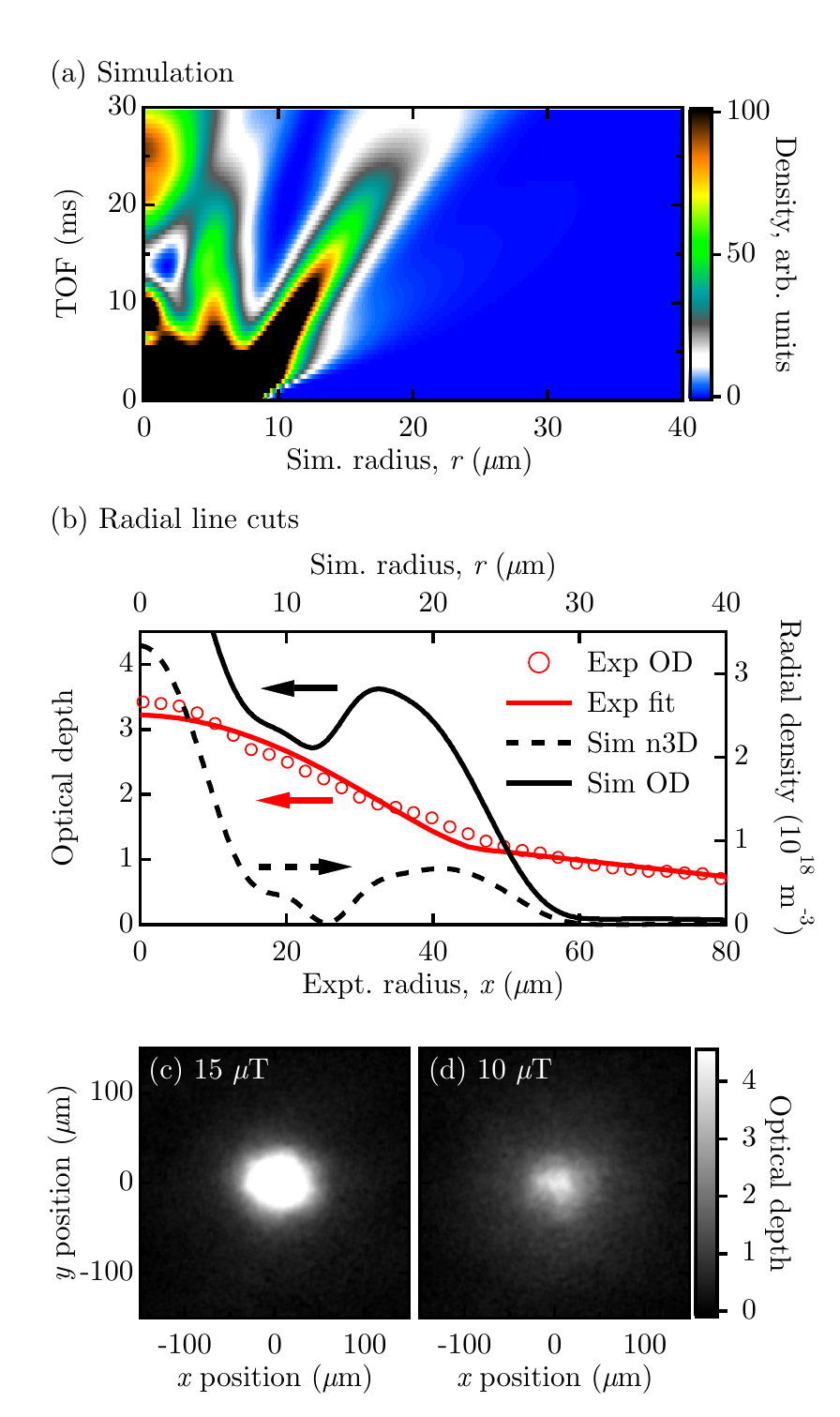}
\caption{
(a) Radial density profile for a simulated cold atom cloud is mapped to a non-linear color scale that saturates at a peak density of $5\times 10^{18}$\ m$^{-3}$.  The density profile evolves as a function of time during TOF.  After an initial expansion, the cloud contracts into concentric shells as the field settles at $10 \mu$T, expanding again after $t=20$~ms when the field is turned off.  (b) A line cut through the simulated image (a) at $t=30$~ms shows a secondary peak in the radial density profile (dotted curve), which is still visible when the density is integrated to simulate optical depth (solid black curve).  A suggestive modulation is visible in the corresponding experimentally observed OD profile (open circles), relative to a Thomas-Fermi plus Gaussian fit (solid red curve), but upper (simulated) and lower (experimental) length scales do not match, and results are ambiguous relative to other artifacts which can occur in absorption images.  (Lower panel) Experimental absorption images showing bimodal density profiles at approximately 15\ $\mu$T (c) and 10\ $\mu$T (d).
}
\label{fig:GPsim}
\end{figure}

To better understand the observed inflection in loss versus field in the negative $a_s$ regime, we modeled our system within the GPE, which has the form of the nonlinear Schr\"{o}dinger equation.  Assuming spherical symmetry, we cast the 3D GPE as an effective 1D radial equation \cite{Ruprecht:PRA1995, Savage:PRA2003},
\begin{align}
	&i \hbar \frac{\partial \psi (r,t)}{\partial t}\!=\!\Bigg\{\!-\!\frac{\hbar^2}{2m}\left(\frac{\partial^2}{\partial r^2} + \frac{2}{r}\frac{\partial}{\partial r}\right) + V(r,t)  \label{GPE} \\
	&+ g(t)|\psi(r,t)|^2 -i\frac{\hbar}{2}\left[K_1 + K_3(t)\left|\psi(r,t)\right|^4\right] \Bigg\} \psi(r,t),
\nonumber
\end{align}
where the time-dependent radial wavefunction $\psi(r,t)$ is normalized to the total atom number $N$.  The potential $V(r,t<0) = m \omega^2 r^2 /2$ while $V(r,t \geq 0) = 0$, corresponding to the release of the atoms from the trap at $t=0$.  The time-dependent interaction strength $g(t)=4\pi\hbar^2 a_s(t)/m$ where $a_s(t<0) = a_{\rm bg}$ and $a_s(t \geq 0)$ is given by Eq.~\eqref{eq:scattering} with the inclusion of a time-dependent magnetic field $B(t)$.  The phenomenological loss term is equivalent to the integral term in Eq.~\eqref{eq:K1andK3}, but is completely general, making no assumptions about the form of the density profile.  The loss constants $K_1$ and $K_3[B(t)]$ are taken from the Castin-Dum analysis of the data of Fig.~\ref{fig:alphaK3}.  The time-dependence of the field is also the same as for the Castin-Dum analysis above.
%

We numerically solved the GPE using the Crank-Nicolson (CN) method \cite{Williams:Thesis1999, Garcia:NumMethods2000}.  Beginning with a trapped condensate ($\omega/ 2 \pi =65$~Hz) at a magnetic field of $B-B_0=0.2$~mT, the ground state was determined by initializing the wave-function with a Thomas-Fermi profile, then propagating in imaginary time to eliminate higher order spatial modes \cite{Williams:Thesis1999}, resulting in a close approximation to the true ground state.  The wavefunction was normalized to $N$ particles after each imaginary time-step, where $N$ ranged from $2.5\times10^4$ to $2.5 \times 10^5$ atoms.  The CN algorithm was then run in real time with the trapping potential turned off; the field was ramped to a final field $B_f$, and the atom number was allowed to diminish according to the loss constants.  After $t=20$~ms, the magnetic field was instantaneously set to 0.  The simulation was then allowed to evolve for an additional 10~ms, corresponding to a total of 30~ms TOF, in order to provide direct comparison with experimental results.  Finally, the radial density profiles were integrated along one dimension so that the resulting profiles correspond to the optical depth (OD) profile that is obtained experimentally from an absorption image.

Fig.~\ref{fig:GPsim}(a) shows the simulated 1D radial density profile evolving during time-of-flight (TOF).  Following release at $t=0$, the BEC containing $2.2 \times 10^5$ atoms undergoes initial expansion for $\approx 1$~ms before the field settles to a value of $B-B_0 = 10~\mu$T, where $a_s=-0.7a_{\rm bg}$. Despite the negative scattering length, corresponding to attractive mean field energy, the condensate does not immediately collapse.  Rather, the kinetic energy imparted to the system during the brief period following release (while the mean field energy was still positive) imposes a quadratic spatial dependence on the phase of the wavefunction, during an initial outward expansion of the condensate.  When $a_s$ becomes negative, the outward velocity of the wavefunction is decelerated by the attractive mean field energy.  However, because the BEC's density is non-uniform, the magnitude of deceleration depends on position.  For sufficiently negative $a_s$, the BEC's outward expansion can reverse, but owing to the inhomogenous density, this reversal need not be complete; the in- and out-going components can interfere.  This explains the arms that begin to form in the 1D radial density map around $t=5$~ms.  Translated into three dimensions, these arms correspond to the development of concentric shells.  At $t=20$~ms, the field is turned off completely, and the scattering length becomes positive once again.  The concentric shells or fringes expand and blur, but are still visible in the simulated image at 30\ ms TOF.  The description in terms of interfering matter waves is appealing, and provides for a simple intuitive understanding of this physics. It is, however, not required: the appearance of density modulations (rings) can also be fully explained using a suitable set of classical fluid hydrodynamic equations, including 3-body recombination.

Fig.~\ref{fig:GPsim}(b) shows a line cut through the radial density image of Fig.~\ref{fig:GPsim}(a) at TOF = 30 ms.  The simulated 1D radial density profile (dotted curve) shows a strong secondary peak corresponding to a shell of atoms at $r \approx 20~\mu$m.  The optical depth profile (solid black curve) is calculated from the 3D density distribution corresponding to the radial density profile, by integrating along a Cartesian axis.  The prominent secondary peak in the radial density profile is still visible in the calculated OD profile, which is what should be observed in experiment.  A line cut from the experimentally observed OD profile (open circles), obtained with $B_f\approx10\ \mu$T, shows a small amount of spatial modulation compared to the Thomas Fermi plus Gaussian fit to the data (solid red curve), but similar artifacts can sometimes be seen far from the Feshbach resonance, and are decidedly inconclusive.   Note that the range of the $x$-scale for the experimental data differs from that of the $r$-scale for the simulation results.  The absence of any clear indication of shell structure in our experimental results may be due to imperfect trap symmetry, the blurring that occurs in the final 10 ms of expansion, a lack of spatial resolution in our imaging setup, or ``seeding" of the gain processes by thermally induced modulations in density.  Rapid heating, which occurs below $B_f\approx 12\ \mu$T, also contributes to the ambiguity of our results, as illustrated by the increased thermal halo in the experimental absorption image of Fig.~\ref{fig:GPsim}(d), obtained with $B_f\approx10\ \mu$T, relative to Fig.~\ref{fig:GPsim}(c), obtained with $B_f\approx15\ \mu$T.

The GPE simulations were performed at several different fields to obtain final cloud characteristics for comparison to Fig.~\ref{fig:numandwidth}.  For small $B-B_0$, the appearance of the secondary fringe presents a challenge to identification of the Thomas-Fermi radius.  Our procedure is to ignore the central peak which can, in simulation, become quite prominent, so that the overall envelope is no longer Thomas-Fermi.  We therefore mask the data to select only the monotonic region of the secondary fringe and fit this to a Thomas-Fermi (inverted parabola) profile.  Fig.~\ref{fig:numandwidth} shows the results of the GPE simulation (dotted curves) for the high side of the resonance.  As for the CD numerical solution, the GPE simulation reproduces the number loss data quite well, including the inflection in the slope of the loss curve.  In Fig.~\ref{fig:numandwidth}(c), the GPE simulation underestimates the magnitude of $R_{\rm TF}$, but qualitatively reproduces the upturn in $R_{\rm TF}$ at low values of $B_f-B_0$.

\begin{figure}
\center
\includegraphics{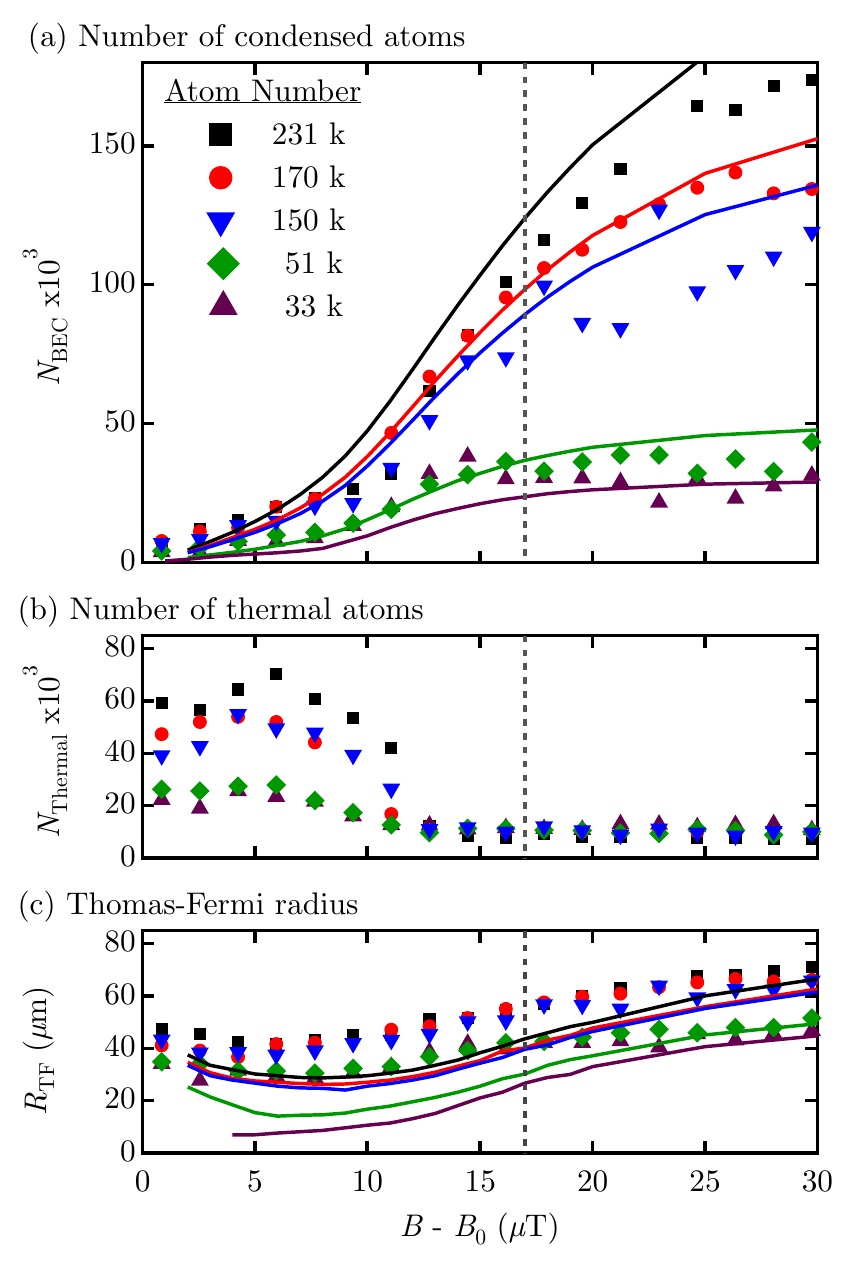}
    \caption{(a) Number of condensed atoms $N_{\rm BEC}$ measured after TOF versus final offset field $B_f-B_0$.  For low initial atom number $N_i$ = 51~k and 33~k atoms, a loss threshold is visible at $B-B_0 \approx 14~\mu$T, significantly less than the theoretical resonance width $\Delta B_{\rm th} = 17~\mu$T, which is indicated by the vertical dotted line.  Solid curves indicate GPE simulation results, which use $\Delta B = 17~\mu$T and which fail to reproduce the observed loss threshold for $N_i$ = 51~k and 33~k atoms.  (b)  Number of thermal atoms $N_{\rm Thermal}$ after TOF.  A threshold for heating is visible at $B_f-B_0 \approx 12~\mu$T.  (c) Thomas-Fermi radius $R_{\rm TF}$ vs final offset field.  The minimum in $R_{\rm TF}$ occurs well below the expected zero crossover for $a_s$, both for experimental data (symbols) and GPE simulations (solid curves). }
    \label{fig:lossvsnum}
\end{figure}

The upturn in $R_{\rm TF}$ observed in our experiment is found, not at $B_f-B_0\approx\Delta B_{\rm th}=17~\mu$T, as previously reported in Ref.~\onlinecite{Volz:PRA2003}, but at a much lower field, $B_f-B_0 < 10~\mu$T.  This difference suggests that the upturn in $R_{\rm TF}$ is a poor measure of resonance width $\Delta B$ which, to our knowledge, has otherwise never been measured.  We therefore consider whether there are any other prominent features in the observed resonance data, particularly near the expected crossover from positive to negative $a_s$.  As shown in Fig.~\ref{fig:lossvsnum}(a), we have measured atom number versus offset field on the high side of the resonance, for several clouds with varying initial atom number.  This is accomplished by preparing the cloud as for the data of Fig.~\ref{fig:numandwidth}, but varying the magneto-optical trap loading time from a few hundred ms to several seconds.  For the three datasets with highest initial atom number $N_i = 2.3 \times 10^5$, $1.7 \times 10^5$, and $1.5 \times 10^5$, the observed loss feature is unchanged in the high side data of Fig.~\ref{fig:numandwidth}(a).  GPE simulations (solid curves) show reasonable qualitative agreement with these data.  For lower $N_i$ = $5.1 \times 10^4$ and $3.3 \times 10^4$, the atom number is nearly independent of final field offset until $B_f-B_0 \approx 14~\mu$T, which appears as a threshold field for atom loss.  The GPE simulations, which assume $\Delta B = 17~\mu$T, do not reproduce this threshold behavior.

The threshold behavior observed in $N_{\rm BEC}$ is accompanied at slightly lower field by an increase in thermal atom number $N_{\rm thermal}$, as shown in Fig.~\ref{fig:lossvsnum}(b).  The increase in thermal number is abrupt for all values of $N_i$, and occurs at a final offset field $B_f-B_0 \approx 12~\mu$T that is significantly lower than the theoretical width $\Delta B_{\rm th} = 17~\mu$T, which is marked by a dotted line in the figure.  The presence of thermal atoms is not included in the GPE simulations, since the 3-body loss mechanism is usually associated with loss of all three atoms from the trap.  On the low side of the resonance, the existence of a weakly bound state could mediate three body recombination at sufficiently low energies that the atoms remain trapped.  On the high side of the resonance, however, no such weakly bound state exists.  

We recall that the shell structure of Fig.~\ref{fig:GPsim} is not definitively observed in our experiment.  However, the formation of this shell structure can be viewed as an interference effect, and self-interference has been associated with turbulence, vortex formation, and heating \cite{Jo:PRL2007}.  It is conceivable that the onset of heating in our experiment is an indication of self-interference of the condensate as it collapses under the influence of attractive interactions.  In simulations, the onset of interference effects, manifest in the formation of a secondary peak in the density profile, occurs within $1~\mu$T to $2~\mu$T of the zero crossing of $a_s$.  We therefore have a second possible indication of the true resonance width.

Finally, we see in Fig.~\ref{fig:lossvsnum}(c) that the increase in $R_{\rm TF}$ at very low values of $B_f-B_0 <10~ \mu$T is a robust phenomenon for any $N_i$.  We interpret this phenomenon in terms of the shell structure that appears in the density distribution of Fig.~\ref{fig:GPsim}.  Interference modifies the usual Thomas-Fermi (inverted parabola) density distribution, such that the contracting cloud of atoms leaves behind a shell of atoms at a radius that is larger than a simple Castin-Dum scaling law would predict.  Here we have to posit a conspiracy of heating and other imperfections that transform this shell of atoms into an effective increase in the overall diameter of the cloud.  This is clearly unsatisfying.  We note that the GP simulations in this paper solve the 3D GPE as an effective 1D radial equation.  It is possible that a full 3D simulation including initial trap assymetries would fail to produce the shell structure of Fig.~\ref{fig:GPsim} or would transform it in some other way.

%
\section{\label{sec:Summary}Summary}
We have shown that the collapse of an untrapped $^{87}$Rb condensate with negative scattering length can be made to proceed in a stable manner.  The condensate is stabilized against explosive collapse in part by the kinetic energy imparted upon its release from the trap.  The observed increase in $R_{\rm TF}$ may be related to a self-interference effect that imposes a shell structure onto the condensate, as seen in GPE simulations.  While this shell structure is not observed directly in experiment, self-interference may be responsible for turbulence resulting in the observed increase in thermal fraction at low offset field.

\begin{acknowledgments}
We thank W.~D. Phillips for discussions.  This work was partially supported by ONR, ARO with funds from the DARPA OLE program, and the NSF through the JQI Physics Frontier Center. R.L.C. acknowledges the NIST/NRC postdoctoral program and K.J.G. thanks CONACYT.
\end{acknowledgments}


\begin{thebibliography}{18}
\bibitem{Current}
Current is provided by a commercial linear power supply capable of supplying up to 400~A at 60~V with a current resolution of 10~ppm.  To achieve field stability $\approx 10$~ppm with a short settling time, the commercial supply is switched between the 48 turn field coils and a matching resistive load that is spatially well separated from the experimental chamber, reducing the settling time to under 1~s.
\end{thebibliography}
\end{document}